\documentclass[10pt,conference]{IEEEtran}
\IEEEoverridecommandlockouts
\usepackage{cite}
\usepackage[normalem]{ulem}
\usepackage{amsmath,amssymb,amsfonts}
\usepackage{algorithmic}
\usepackage{graphicx}
\usepackage{textcomp}
\usepackage{xcolor}
\def\BibTeX{{\rm B\kern-.05em{\sc i\kern-.025em b}\kern-.08em
    T\kern-.1667em\lower.7ex\hbox{E}\kern-.125emX}}
\begin{document}

\title{Detecting User-Perceived Failure in Mobile Applications via Mining User Traces}

\author{\IEEEauthorblockN{Deyu Tian}
\IEEEauthorblockA{\textit{Key Lab of High-Confidence Software Technology (MoE), Peking University}\\
tiandeyu@pku.edu.cn}
}

\maketitle

\begin{abstract}
Mobile applications (apps) often suffer from failure nowadays. Developers usually pay more attention to the failure that is perceived by users and compromises the user experience. Existing approaches focus on mining large volume logs to detect failure, however, to our best knowledge, there is no approach focusing on detecting whether users have actually perceived failure, which directly influence the user experience. In this paper, we propose a novel approach to detecting user-perceived failure in mobile apps. By leveraging the frontend user traces, our approach first builds an app page model, and applies an unsupervised detection algorithm to detect whether a user has perceived failure. Our insight behind the algorithm is that when user-perceived failure occurs on an app page, the users will backtrack and revisit the certain page to retry. Preliminary evaluation results show that our approach can achieve good detection performance on a dataset collected from real world users.
\end{abstract}


\section{Introduction}\label{sec:introduction}
Mobile devices are continuously gaining their popularity in recent years.
With the help of mobile apps, users can chat with friends, buy things in e-shop, book hotel rooms, and buy flight tickets, totally online. Mobile apps have greatly reshaped people's daily lives.

However, a user may perceive failure that hinders her further progress, when she tries to use a functionality provided by the mobile app but fails.
Failure in mobile apps will reduce the productivity of their users, which harms the user experience of the mobile app, leading to user churn and revenue loss \cite{app_crash_hurting_revenue}.
So, it is urgent for developers to find out whether users have perceived failure and fix the potential defects in mobile apps as soon as possible.

However, there might be a gap between the user-perceived failure and the failure detected by existing approaches.
These approaches, ranging from using simple heuristics to detect failure symptoms to adopting complex machine learning system to detect anomalies \cite{Du2017deeplog}, detect failure from the perspective of software logs instead of user behavior or user experience.
In reality, however, some failure may be already handled by fault-tolerance mechanism \cite{Inukollu2015design}, but still leaves warning or error events in the logs. Some failure has been perceived by the users, but may still be omitted by mining software logs.
These factors make a gap between the existing approaches and the user experience.

So in this paper, we propose a new complementary approach to addressing the problems that exist in current practice. Our approach is based on analyzing user behavior instead of software logs. We leverage a large quantity of navigation traces in a mobile app that are collected from real users, to detect whether users have perceived failure while using the mobile app.
Even though user traces are simple and easily available data, our approach can still find a lot of perceived failure.
Meanwhile, our approach respects user privacy; sensitive data like the parameters of a page and the user interactions inside a page are not collected in user traces. 

The insight behind our approach is that when a user perceives failure in a mobile app, the user will backtrack from the failure page, and try to solve the issue by revisiting the page and checking whether the same issue still exists.
So our detection approach is based on two features in user traces: user backtracking and the extent of it.
To this end, we leverage a Markov model and design an algorithm to distinguish whether users will backtrack from some app pages normally.
Then, we propose an unsupervised anomaly detection algorithm, to detect users that probably have perceived failure.

We evaluate our approach on a real world dataset, and the preliminary results show that our approach can detect user-perceived failure issues well.

\section{Background \& Related Work}\label{sec:background}
If a user cannot make progress in a task that is covered by the functionalities of the mobile app due to some technical faults of the app, we say the user {\em perceives failure}.
Meanwhile, we call the app page where the perceived failure occurs, as the {\em failure page}.

A sequence of user navigation events constitutes a {\em user trace}. A navigation event contains the user ID, the page ID of the target app page, and the time stamp when the navigation is triggered, excluding the {\em page parameters}.
These parameters usually contain highly sensitive user data, so we do not collect these parameters for respecting user privacy.

Users usually will navigate between different app pages to accomplish a task.
So for a specific task, different app pages have different closeness to the completion of the task, which we will call as {\em progress value} of an app page.
If a user navigates from a page with high progress value to a page with low progress value, we will call such navigation as {\em backtracking}. In our work, user backtracking is an important indicator of whether a user has perceived failure.

There has been mature support for collecting user traces in mobile apps \cite{google_analytics_path_analysis}, which makes user traces an easily available and widely used source of data. It is common practice in both research and industry \cite{industry_user_behavior_analytics} to understand user behaviors by analyzing user traces.
Related research topics include user behavior analysis \cite{Benevenuto2009characterizing} \cite{Wang2017clickstream} \cite{Liu2019characterizing} \cite{Sinha2014your}, malicious behavior identification \cite{Wang2017clickstream} \cite{Shi2019detecting} \cite{Obendorf2007web}, user trace visualization \cite{Mysore2018porta} \cite{Zhao2015matrixwave}, and advertisement \cite{Wang2019quality}. In this paper, our approach can detect user perceived failure while leverages simpler user trace data than prior work, for respecting the user privacy.

Log analysis \cite{He2020survey} is heavily adopted in anomaly detection of software behavior.
Researchers have proposed multiple approaches to detect anomalies in software logs \cite{He2016experience}, including PCA \cite{Xu2009detecting}, clustering \cite{Lin2016log} \cite{He2018identifying}, SVM \cite{Liang2007failure}, frequent pattern mining \cite{Farshchi2015experience} \cite{Shang2013assisting}, random forest \cite{Zhou2019latent}, word2vec \cite{Bertero2017experience}, automaton \cite{Debnath2018loglens} \cite{Amar2018using} \cite{Yu2016cloudseer} \cite{Beschastnikh2014inferring}, and  deep learning \cite{Du2017deeplog} \cite{Meng2019loganomaly} \cite{Liu2019log2vec} \cite{Zhou2019latent}. These approaches mainly focus on traditional software logs or server logs, and are not designed to analyze user traces and detect the user-perceived failure.

\section{Methodology}\label{sec:methodology}
\begin{figure}
    \centering
    \includegraphics[width=0.5\textwidth]{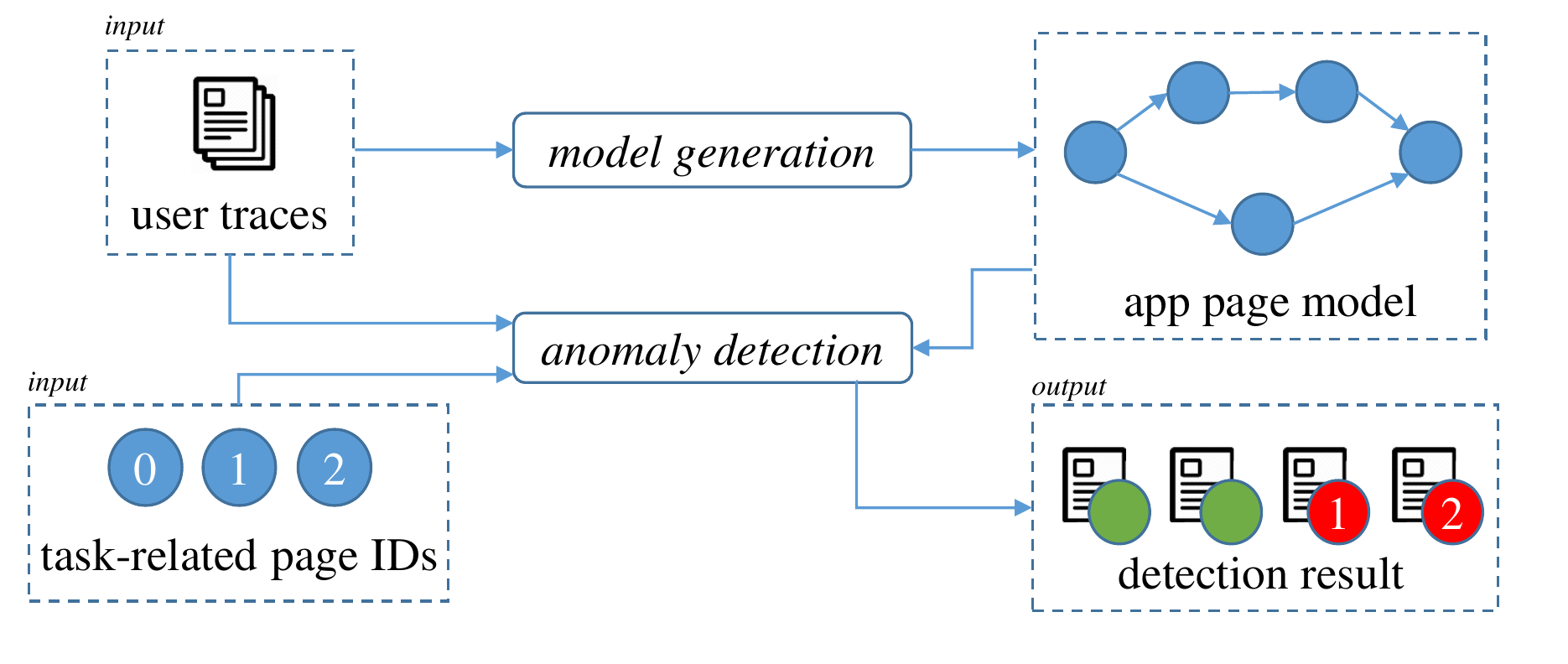}
    \caption{Overview of our approach}~\label{fig:overview}
    \vspace{-1em}
\end{figure}
The overall methodology overview is shown in Figure \ref{fig:overview}. 
Our approach essentially has two steps: model generation, and anomaly detection.
Ahead of that, a step of data preparation is necessary before developers apply our approach.

{\em Data preparation.} First developers should prepare the data to feed in our approach.
Our approach concentrates on one task in a mobile app at each time.
The developers should choose a task and a date, and sample a set of users that have tried to use that task in the day. To specify a task, developers should specify all the pages that are involved in the task, together with the beginning page and the final page of the task.

{\em Model generation.} In this step, we will generate an app model from the user traces. The app model is a Markov model \cite{Sadagopan2008characterizing}, in which each node represents an app page and each edge represents the navigation between the two pages. Each node in the model has a progress property, which denotes the progress value of the page; each edge in the model has a probability property, which denotes the probability that the user will navigate to the target page if the user is at the source page.
The progress value of an app page can well reflect how much progress that a user has made with regards to the task when the user navigates to the app page.

{\em Anomaly detection.} In this step, we will detect whether a user has perceived failure, based on the model that we build before.
Our insight is that a user will backtrack an abnormal large quantity of times when the user perceives failure.
First, we leverage the Markov model, to derive a mathematical probability distribution about the times that an average user will backtrack from an app page.
Based on the probability model, we can estimate the effectiveness of the detection on some app pages. If users will backtrack from an app page a lot of times normally, we will exclude the app page from later anomaly detection.

In anomaly detection, the core pattern that we focus on, is that a user advances from page A to page B, and then backtracks from page B to page C (A and C can be the same page).
We will extract two features from a user trace and a task-related app page: (a) the number of this pattern (b) the intensity of this pattern. To get the number of this pattern, we simply count how many times this pattern appears in each user trace. To get the intensity of this pattern, we count the maximum consecutive times that this pattern appears while the time interval between two adjacent appearances of this pattern is under 2 minutes.
After extracting the two features, we can compute an anomaly score for each user trace.
First, we focus on each feature, and find its distribution among all user traces and get the cumulative distribution function $CDF(x)$. Then we compute the anomaly score of a user trace with feature value $v$ as $1-CDF(v)$.
Finally, we sort the user traces according to their anomaly scores, and predict the user traces as anomalies if their scores exceed a predefined threshold $\epsilon$.

\section{Preliminary Results \& Discussion}\label{sec:evaluation}
We have conducted a preliminary experiment to evaluate the performance of our approach. 
We cooperate with a mobile app company, and focus on a single mobile app functionality where users will surely perceive failure if failure occurs.
We pick a random day, and collect a dataset that includes 5183 traces of the users who have used the functionality in that day.
Meanwhile, we manually design some heuristic rules based on the server logs to label whether failure occurred in a user trace.
Finally in the dataset, we find that 1619 users should have perceived failure.

We evaluate our approach on the dataset. We set the threshold $\epsilon$ as 0.8, and the result shows that our approach achieves 74\% precision and 75\% recall in detecting user-perceived failure.

Currently we evaluate our approach in only one dataset. In future work, we will evaluate our approach and other related baseline approaches in more datasets as well as improving the performance of our approach.

\section{Conclusion}\label{sec:conclusion}
In this paper, we design a novel approach, to detect user-perceived failure in mobile applications by a large quantity of user traces. We build a page transition model of a mobile app, and detect anomalous user traces according to the insight that users will backtrack and revisit the failure page. The preliminary evaluation results show that our approach is effective at detecting user-perceived failure.

\bibliographystyle{IEEEtran}
\bibliography{related}

\end{document}